\documentclass{article}

\usepackage{arxiv}

\usepackage[utf8]{inputenc} 
\usepackage[T1]{fontenc}    
\usepackage{hyperref}       
\usepackage{url}            
\usepackage{booktabs}       
\usepackage{amsfonts}       
\usepackage{nicefrac}       
\usepackage{microtype}      
\usepackage{cleveref}       
\usepackage{lipsum}         
\usepackage{graphicx}
\usepackage{natbib}
\usepackage{doi}
\usepackage{indentfirst}

\title{Physics-based model of solar wind stream interaction regions: Interfacing between Multi-VP and 1D MHD for operational forecasting at L1}

\author{ 
R. Kieokaew\thanks{Institut de Recherche en Astrophysique et Planétologie, UPS, CNES, CNRS, Toulouse, France \texttt{(rkieokaew@irap.omp.eu)}} \and
R. F. Pinto \thanks{Departement d’Astrophysique/AIM, CEA/IRFU, CNRS/INSU, Univ. Paris-Saclay \& Univ. de Paris, Gif-sur-Yvette, France} \footnotemark[1] \and
E. Samara \thanks{CmPA, KU Leuven, Leuven, Belgium} \and
C. Tao \thanks{National Institute of Information and Communications Technology (NICT), Koganei, Tokyo, Japan} \and
M. Indurain\footnotemark[1] \and
B. Lavraud \thanks{Laboratoire d’astrophysique de Bordeaux, Univ. Bordeaux, CNRS, Pessac, France} \footnotemark[1] \and 
A. Brunet \thanks{ONERA/DPHY, Université de Toulouse, F-31055 Toulouse, France} \and 
V. Génot  \footnotemark[1] \and
A. Rouillard \footnotemark[1] \and
N. André \footnotemark[1] \and 
S. Bourdarie \footnotemark[7] \and 
C. Katsavrias \thanks{Department of Physics, National and Kapodistrian University of Athens, 15784 Athens, Greece} \and
F. Darrouzet \thanks{Royal Belgian Institute for Space Aeronomy (IASB-BIRA), Brussels, Belgium} \and
B. Grison \thanks{Institute of Atmospheric Physics CAS, Dept of Space Physics, Prague, Czech Republic} \and
I. Daglis \footnotemark[8] 
}

\renewcommand{\shorttitle}{Automated CIR pipeline}

\hypersetup{
pdftitle={Automated CIR pipeline},
pdfsubject={space weather},
pdfauthor={R. Kieokaew},
pdfkeywords={Space weather, solar wind, corotating interaction region},
}

\begin{document}
\maketitle

\begin{abstract}{
Our current capability of space weather prediction in the Earth's radiation belts is limited to only an hour in advance using the real-time solar wind monitoring at the Lagrangian L1 point. To mitigate the impacts of space weather on telecommunication satellites, several frameworks were proposed to advance the lead time of the prediction. We develop a prototype pipeline called ``Helio1D'' to forecast ambient solar wind conditions (speed, density, temperature, tangential magnetic field) at L1 with a lead time of 4 days. This pipeline predicts Corotating Interaction Regions (CIRs) and high-speed streams that can increase high-energy fluxes in the radiation belts. The Helio1D pipeline connects the Multi-VP model, which provides real-time solar wind emergence at $0.14$ AU, and a 1D MHD model of solar wind propagation. We benchmark the Helio1D pipeline for solar wind speed against observations for the intervals in 2004 - 2013 and 2017 - 2018. We developed a framework based on the Fast Dynamic Time Warping technique that allows us to continuously compare time-series outputs containing CIRs to observations to measure the pipeline's performance. In particular, we use this framework to calibrate and improve the pipeline's performance for operational forecasting. To provide timing and magnitude uncertainties, we model several solar wind conditions in parallel, for a total of 21 profiles corresponding to the various virtual targets including the Earth. This pipeline can be used to feed real-time, daily solar wind forecasting that aims to predict the dynamics of the inner magnetosphere and the radiation belts.}
\end{abstract}


\section{Introduction}
The Sun and the solar wind are the main drivers for space weather at Earth and in the heliosphere. Nowadays, forecasting solar wind conditions becomes a crucial task as the human society is increasingly dependent on space technologies that are under the influence of space weather. The major drivers that can have profound effects are the Coronal Mass Ejections (CMEs), caused by solar eruptions, and Corotating Interaction Regions (CIRs), formed when a fast solar wind or high-speed stream (HSS) from the coronal holes takes over a slower solar wind. Although CMEs cause more severe events, their effects often disappear in a day or two while CIR/HSS events have significant effects that last several days after their arrivals \citep[e.g.][]{2012JASTP..83...79B}. During solar minima when CMEs are absent, CIRs cause low- to intermediate-strength geomagnetic storms and occasionally lead to strong geomagnetic storms \citep[e.g.][]{2001GL013052, 2006JGRA..111.7S05A, 2006JGRA..111.7S01T, 2008JGRA..113.8106Z, Kilpua:2017aa, 2018SW001894}. These CIR-driven storms also have impacts down to the ionosphere and thermosphere, leading to equatorial ionization anomaly \citep[e.g.][]{2022AdSpR..69.2168O} and total electron content variability \citep[e.g.][]{2020AdSpR..65..198C}. Furthermore, CIRs impact the dynamics of CMEs \citep[e.g.][]{2019ApJ...887..150L}, broaden solar energetic particle spectrum \citep[e.g.][]{2016JGRA..121...77Z, 2019A&A...624A..47W}, and modulate galactic cosmic rays \citep[e.g.][]{1979ICRC....3..438I, 2007SoPh..245..191R}. \citet{2017SSRv..212.1345C, Kilpua:2017aa, 2017SoPh..292..140V, 2018LRSP...15....1R} and \citet{2018JASTP.180...52Y} comprehensively reviewed our current physical understanding and the geoeffectiveness of CIRs. 

Communication satellites in the geostationary orbit and medium and low Earth orbits are in the vicinity of the Earth's radiation belts - the regions encircling the near-Earth space and containing significant fluxes of high-energy electrons and ions. CIRs can drive radiation belt electron variability \citep[e.g.][]{2009JASTP..71.1145L, 2021SpWea..1902882H}, leading to electron flux enhancements \citep[e.g.][]{1997GeoRL..24..927B, 1997GeoRL..24..923L}; electron acceleration especially in the presence of strong southward Interplanetary Magnetic Field (IMF Bz) component \citep{1997GeoRL..24..927B}, in which in certain cases reaching relativistic level \citep{1979GMS....21..180P, 2018JGRA..123.1134P, 2021SpWea..1902882H, https://doi.org/10.1029/2019JA026771}. The high speed streams that follow CIRs have been demonstrated to be more effective (compared to CMEs) in producing multi-MeV electron enhancements up to more than 7 MeV \citep{2018SpWea..16.1202H, https://doi.org/10.1029/2019JA026743}. This is due to fact that they produce intervals of combined southward IMF and solar wind velocity over 500 km/s, which lead to an enhanced magnetic reconnection rate at the dayside magnetopause \citep{https://doi.org/10.1029/2005JA011447, https://doi.org/10.1002/grl.50916}. 

Since the beginning of the space technology era, there have been a number of reports of spacecraft anomalies and even failures, for example, at the geostationary orbit due to the elevated fluxes of several MeV electrons that persisted for several days following CIRs \citep[see a review by][]{Lanzerotti2007}. To better mitigate impacts on those satellites, we need to develop a system for continuously predicting the solar wind and radiation belt conditions. With the real-time monitoring of the solar wind at the Lagrangian L1 point, e.g. by  Advanced Composition Explorer (ACE), the continuous, real-time prediction of the radiation belt conditions via modeling is limited to one hour in advance or less as the solar wind takes about 40 - 60 minutes from the ACE spacecraft to arrive at the magnetosphere. Using the Kp that is a 3-hour index indicative of global auroral activities, the SPACECAST \citep{2013JSWSC...3A..20H} model, for instance, was used to provide a forecast up to 3 hours in advance based on the real-time ACE data. To further improve our current capability, we need continuous solar wind prediction with a significant lead time. 

Numerous efforts have been made in modeling real-time solar wind conditions at Earth. The majority of models consists of two main parts: (1) a young solar wind emergence modeling and (2) a solar wind modeling and/or propagation from near Sun to Earth. The former part is initiated by a reconstruction of the coronal magnetic fields based on synoptic or syncronic maps from ground observations; this is achieved by employing a simplified analytical model with a magnetostatic potential field source surface (PFSS) extrapolation. The PFSS extrapolation combined with the Wang-Sheeley-Arge (WSA) model is a widely-used predictor of the solar wind in the corona/low heliosphere \citep{2000JGR...10510465A, 2004JASTP..66.1295A}. In combination with the global 3-D magnetohydrodynamics (MHD) ``ENLIL'' --- the physics-based solar wind model \citep{1996JGR...10119973O, 1999JGR...10428225O, 2004JASTP..66.1311O} --- the WSA-ENLIL has been used to provide daily forecasts of solar-wind streams. The WSA-ENLIL tool has also been extended to include CMEs using an ad-hoc hydrodynamic cone model or the spheromak model \citep{2009SpWea...7.3004T, 2011SpWea...9.6002T, 2018shin.confE.120T}. The Space Weather Prediction Centre of the National Oceanic and Atmospheric Administration (SWPC/NOAA) has put the WSA-ENLIL into operation to provide daily solar wind predictions with a lead time of 1 - 4 days \citep{2011SpWea...9.3004P}. Alternative to the WSA model is the MHD Around a Sphere (MAS) that is a global, time-dependent model based on the resistive MHD equations \citep{1999JGR...104.9809L, 2012JASTP..83....1R}. The WSA-ENLIL and the MAS-ENLIL tools have been extensively benchmarked by \citet{2005JGRA..11012105O, 2009SoPh..254..155L, 2009SpWea...7.6004M, 2009SpWea...712002M, 2011SoPh..273..179J, 2011SoPh..274..321J, 2015SpWea..13..316J}. For the solar-wind propagation part, several less-numerically intensive options are also available, though they are not fully put into operation at the time of this writing. A few models include Heliospheric Upwind eXtrapolation (HUX) \citep{2011SoPh..270..575R}, ballistic propagation \citep{Dosa:2018aa}, and Tunable HUX \citep{2020ApJ...891..165R}. \citet{https://doi.org/10.1029/2018SW002040} and \citet{2019ApJS..240...35R, 2020ApJ...891..165R} compiled several combinations of the existing models and established frameworks for cross-validation.   

Recently, more efforts have been made to combine and benchmark European-based models to develop operational solar-wind forecasting services. The Multi-VP is a coronal model developed by \citet{2017ApJ...838...89P} that computes multiple solutions of 1D solar wind flux tubes in sub-domains of interest. This approach allows a much faster calculation compared to global models with better accuracy at regions of interest, e.g. at the sub-Earth point, with a lead time of 1 -- 3 days \citep{2020A&A...642A...2R}. The Multi-VP model has been coupled with EUHFORIA \citep[European Heliospheric Forecasting Information Asset;][]{2018JSWSC...8A..35P} and validated for HSS modeling in comparison with WSA-EUHFORIA \citep{2021A&A...648A..35S}. In addition, the overall performance of solar wind modeling with EUHFORIA has been assessed by \citet{2019SoPh..294..170H} and \citet{Samara2022}. Most recently, the Heliocast --- a global MHD model comprising both coronal and heliospheric parts --- has been developed by Réville et~al. (under revision in JSWSC at the time of this writing) to provide a daily solar wind forecast with a lead time up to 5 days for the French organization for applied research in space weather\footnote{OFRAME, see http://www.meteo-espace.fr/.}. Moreover, an empirical solar wind forecast \citep[ESWF;][]{2007SoPh..240..315V, Milosic2022} model for CIR/HSS prediction with a lead time of 4 days has been developed using an empirical relation found between the areas of coronal holes as observed in Extreme Ultraviolet data and the solar wind speed at 1 AU \citep{2006SoPh..233..265R, 2007SoPh..240..315V}. This model has been put into operation by the European Space Agency.

In this work, we consider the coupling of the Multi-VP model and a 1D MHD model \citep{2005JGRA..11011208T} for the first time to develop an automated pipeline named ``Helio1D'' for solar wind CIR forecasting. The 1D MHD model was originally developed for modeling solar wind conditions at Jupiter using the in situ observations at Earth as inner boundary; it was later extended to provide operational service for solar wind conditions at other solar-system planets \citep{ANDRE201850}, see HelioPropa\footnote{http://heliopropa.irap.omp.eu/}. Here, we model multiple solar wind solutions at several targets around the sub-Earth point in parallel. In essence, we develop a pipeline for daily solar wind forecasting at Earth 4 days ahead with a systematic characterization of timing and amplitude uncertainties. Using 13 years of data covering parts the solar cycles 23 and 24, from 2004 to 2013 and from 2017 to 2018, we benchmark the pipeline results as well as calibrate it to develop a prototype for operational daily solar wind forecasting. The pipeline is integrated into the operational forecasting service prototype on radiation belt environmental indicators for the safety of space assets as a part of the European Union Horizon 2020 SafeSpace\footnote{https://www.safespace-h2020.eu/} project. 

The evaluation or validation of modeling results is crucial to understand the efficiency of the forecast. Several works considered standard metrics such as root mean square, mean square error, and Pearson correlation coefficient to compare modeling results to the observations. With such metrics, a skill score is also computed to assess whether the model performs better than a baseline model. For CIR/HSS and CME modeling, some works also considered event-based verification that allow the characterization of a true positive (predicted and observed), a false positive (predicted but not observed), a false negative (not predicted but observed) to construct contingency tables \citep{woodcock1976evaluation} and compute the probability of detection \citep[e.g.][]{2016SpWea..14..495R}. Most recently, an alternative approach using the Dynamic Time Warping (DTW) algorithm, a technique commonly employed in time-series analysis, was applied to characterize the performance of solar wind modeling \citep{Samara2022}. We consider these metrics and approach for verification of the pipeline's results.  

The paper is organized as follows. Firstly, we introduce the Multi-VP and 1D MHD models and then describe the interfacing between them for the Helio1D pipeline. Secondly, we describe metrics and techniques for quantifying the performance of Helio1D. Thirdly, we provide results from the pipeline with the assessments of the results using the classic metrics and the DTW algorithm. Fourthly, we discuss the pipeline calibration with a newly developed approach with the DTW algorithm. Finally, we discuss operational forecasting and provide conclusions and prospects.

\section{Automated CIR forecasting pipeline}

We develop a pipeline to connect the solar wind forecasting from near-Sun to Earth. Here, we briefly introduce the models for the solar wind formation (Multi-VP) and solar wind propagation (1D MHD). We then describe the interfacing of the two models for the automated Helio1D pipeline.

\subsection{Multi-VP} \label{subsec:mvp}
MULTI-VP \citep{2017ApJ...838...89P} is a coronal model that covers the heliocentric distances over which all solar wind streams are formed and accelerated: between 1 and about 30 solar radii ($R_\odot$). The main advantages are that it is data-driven, while taking into account the thermodynamics of the wind flows across the highly stratified low solar atmosphere, and hence it produces physically correct and realistic estimations of the state of the solar wind across its domain. MULTI-VP determines a full set of physical quantities consisting of solar-wind speed ($\mathbf{V}$), density ($n$), temperature ($T$), magnetic field ($\mathbf{B}$), and secondary quantities without requiring empirical scaling laws. In addition, MULTI-VP actually computes a set of individual wind streams from surface to high corona that can then be put together to build the full three-dimensional solar wind solutions from multiple 1D solar-wind flux tubes. This brings an enormous advantage with respect to more traditional 3D MHD models in terms of required computing time, and also by not being subject to the strong diffusion of the gradients in the transverse directions. In practice, it also lets us compute solar wind solutions restricted to certain regions of interest rather than in the full solar atmosphere. 

For the 1D MHD model, MULTI-VP is required to produce time-series of the solar wind properties at the sub-Earth point at 0.14 AU (at 30 $R_\odot$). Internally, this output is defined via a data object that lists which magnetic field lines (and hence individual wind streams) are sampled, as well as their coordinates and geometry, and properties of the input magnetogram. Each time a forecast is performed, we select a number of positions around the sub-Earth point (i.e. at 0.14 AU in Sun-Earth line) stretching up to 15 degrees in azimuth and 15 degrees in latitude. The spread in latitude is aimed to cover positional errors propagated from the magnetic field reconstruction. The spread in longitude is aimed to cover the azimuthal range that corresponds to an elemental time-series three days of solar rotation with respect to Earth, which translates to one day behind and two days ahead of the modeling at sub-Earth point. The next forecast proceeds in the same way, producing another elemental time-series that partially overlaps the previous in time. This procedure is repeated to continuously build a long-term time-series that can be used as an input to the 1D MHD model.

\subsubsection{Multi-VP data} \label{subsubsec:mvp-data}

In order to characterize the pipeline performance with regard to various phases of the solar cycle, we need sufficiently long datasets. The longest historical magnetogram series available is that from the Wilcox Solar Observatory (WSO). Here, we obtain mainly two long datasets from Multi-VP processed using the WSO magnetograms. The first set extends from Carrington Rotation (CR) 2024 to CR 2139, corresponding to data from December 2004 to August 2013. This 8 year and 7 month-long interval covers the declining phase (2004 – 2008) and the solar minimum (2009 – 2010) of solar cycle 23 as well as the ascending phase (2011 – 2013) of solar cycle 24. The second set extends from CR 2192 to CR 2210 and corresponds to data from June 2017 to November 2018. These two datasets are fed into the 1D MHD model. 
 
\subsection{1D MHD model} \label{subsec:1d}

From $0.14$ AU (30 $R_\odot$) onwards, the solar wind is propagated through the heliosphere using a 1D MHD model \citep{2005JGRA..11011208T} that takes the solar wind time-series as time-varying boundary conditions, the solar wind parameters --- mainly the radial plasma velocity and tangential velocity --- vary as a function of time (see more below). The 1D MHD model propagates the solar wind using an ideal MHD fluid approximation to the target position while taking into account the interaction between fast and slow streams in the radial direction. The code solves the ideal MHD equations under the influence of solar gravity in a one-dimensional spherically symmetric coordinate system. The 1D MHD equations are solved using the Coordinated Astronomical Numerical Software (CANS)\footnote{developed by T. Yokoyama at the University of Tokyo and collaborators, see documentation at http://www.astro.phys.s.chiba-u.ac.jp/netlab/pub/index.html.}, an early version of CANS+ code \citep{2019PASJ...71...83M}. The 1D MHD code was developed by \citet{2005JGRA..11011208T} to propagate the solar wind observations at Earth to upstream of Jupiter, and was further extended by the French plasma physics data center\footnote{Centre de Données de Physique des Plasmas, France. See http://heliopropa.irap.omp.eu/.} to cover other solar-system planets or target spacecraft positions \citep{ANDRE201850}. The code is robust and widely used for solar wind propagation and CIR modeling in the heliosphere \citep[e.g.][]{2021JGRA..12629770P, Nilsson2022}. 

The MVP output at 30 $R_\odot$ were used as boundary conditions for the initiation of the 1D-MHD code, which then propagated these results up to the L1 point at 1 AU. The coordinate system in the 1D MHD code is equivalent to the Radial-Tangential-Normal (RTN) system, where the $X$-axis ($R$) is pointing radially outward from the Sun in the equatorial plane to the Earth, the $Z$-axis ($N$) is the solar rotation axis, and the $Y$-axis ($T$) completes the orthonormal system. The outer boundary is set to 1.4 AU where the derivatives of all physical parameters diminish. The grid spacing and the time step are chosen to be $(1.4 - 0.14)/400$ AU and $150$ s, respectively. To adopt the 1D simulation, we keep the tangential magnetic field $B_y$ component to physical values while the $B_x$ is fixed to 0.001 nT to meet the solenoidal criterion, and the $B_z$ is set to zero (see \citet{2005JGRA..11011208T} for discussion). We retrieve Multi-VP data at 30 minute cadence and resample them to 1 hour cadence; these hourly-averaged data are then linearly interpolated to meet the CFL (Courant-Friedrichs-Lewy) condition for every time step. 

\subsection{Helio1D - Interfacing Multi-VP and 1D MHD} \label{subsec:interfacing}

To automate the interfacing between the two models, several steps are taken. Multi-VP provides solar wind time series with a full set of physical parameters including plasma number density, velocity, temperature, magnetic field in the RTN coordinates. We first format the time series to fit the input format of the 1D MHD model. Here, we take the input at a 1-hour time resolution and output the propagated time series at L1 with the same 1-hr resolution. To successfully run the 1D MHD code, two criteria are required to be met: (1) the data length must cover at least a solar rotation, i.e., 27.25 days, in order to produce a consistent CIR formation, and (2) the parameter values must be physical (e.g. non-negative plasma number density). In certain conditions, the Multi-VP model may provide unphysical solar wind values due to the poor quality of the magnetograms. We apply a set of criteria to remove unphysical values (see Appendix) and fill gaps using linear interpolation between two available data points. These aspects are further described in Section~\ref{sec:operation}.

\begin{figure}
\centering
\includegraphics[width=\columnwidth]{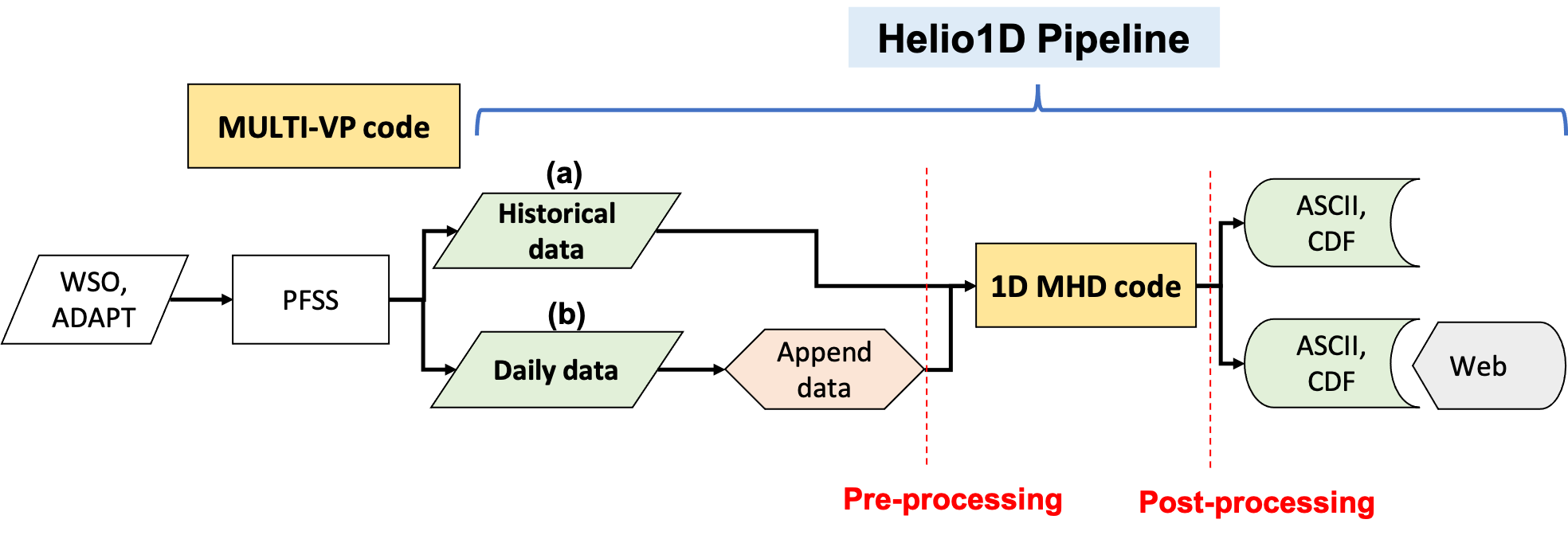}
\caption{Diagram of the Helio1D pipeline.} 
\label{fig:helio1d-pipeline}
\end{figure}

We set up the Helio1D pipeline with two modes for running (a) long-term historical data consisting of several Carrington rotations and (b) short-term data of 1-month long for daily, operational forecasting as shown in Fig~\ref{fig:helio1d-pipeline}. The mode-(a) allows us to quantify the performance of the pipeline as well as identify calibrations that may improve the pipeline performance for operational forecasting. We first benchmark the pipeline using long-term Multi-VP data in Section~\ref{subsubsec:int-results}. 

\section{Measuring the performance of Helio1D} \label{sec:perfm}

In this section, we first introduce the metrics and methodology for characterizing the pipeline performance. We then give results and discuss calibration for the pipeline. 

\subsection{Standard metrics}

We choose three standard metrics - root mean square error (RMSE), mean square error (MAE), and Pearson correlation coefficient ($r$) - to measure the performance of the pipeline. RMSE is a classic point-by-point metric that compares the modeled value and observed value for a given point. It is defined as 
\begin{equation}
RMSE = \sqrt{\frac{1}{T} \Sigma_{t=1}^T \left[ V_m (t) - V_o(t) \right]^2 }
\end{equation}
where $V_m (t)$ and $V_o (t)$ are the modeled and observed values, respectively, at a given time $t$ while $T$ is the number of time elements. Mean absolute error (MAE) is an alternative metric that measures a similar property. It is defined as
\begin{equation}
MAE = \frac{1}{T} \Sigma_{t=1}^T | V_m (t) - V_o(t) |
\end{equation}
where the symbols have the same meanings as defined for the RMSE. The units of the RMSE and MAE are the same as the unit of parameter values $V_m (t)$ and $V_o (t)$. Smaller values of RMSE and MAE indicate better model prediction when compared to observations. Zero values of both metrics indicate a perfect model prediction. 

We also introduce the Pearson correlation coefficient (Pcc) to roughly quantify the similarities between the two time series. In principal, Pearson correlation measures the linear relationship between two datasets. Using the observed and modeled time series ($V_o (t)$ and $V_m (t)$, respectively), the formula for $Pcc$ is 
\begin{equation}
Pcc = \frac{ \Sigma_{t=1}^T  \left( V_m (t) - \langle V_m (t) \rangle \right) \left( V_o (t) - \langle V_o (t) \rangle \right)}{\sqrt{ \Sigma_{t=1}^T \left( V_m (t) - \langle V_m (t) \rangle \right)^2}\sqrt{ \Sigma_{t=1}^T \left( V_o (t) - \langle V_o (t) \rangle \right)^2}}.
\end{equation} 
The $Pcc$-value ranges from $-1.0$ to $1.0$, with a value of $1.0$ being a perfect positive linear correlation between the two time series and $-1.0$ being a perfect negative linear correlation between the two.  

Additionally, to evaluate the model performance in a more concrete way using above metrics, we introduce the skill of a forecast as
\begin{equation}
Skill = 1 - \frac{MSE_{pred}}{MSE_{ref}} \label{eq:skill}
\end{equation}
where $MSE_{pred}$ is the mean squared error of the prediction compared to the observation and $MSE_{ref}$ is the mean squared error of the prediction compared to a reference baseline model. Here, we will use the average of the observation as a reference baseline model (i.e., the climatological mean in climate studies). The observations of solar wind data at L1 are taken from the High Resolution OMNIweb database \citep{2005JGRA..110.2104K}. The average of the OMNI data of the corresponding interval are taken as a reference baseline model. Skill score equal to unity means a perfect forecast while a zero skill means that the model performs the same as the baseline model. On the other hand, a negative skill means that the model performs worse than the baseline model. The aforementioned metrics will be used throughout this report for the quantification of the performance of the model.

\subsection{Dynamic Time Warping Technique}

Although the classic metrics --- RMSE, MAE, skill score, and $Pcc$ --- provide quantification of the performance of the model, they do not provide in-depth information such as time lags, i.e., the difference in arrival times between the modeled and observed stream interfaces, that are crucial for model evaluation. For this reason, we also consider an alternative technique called Dynamic Time Warping (DTW). The DTW technique compares two sequences by finding an optimal alignment by which one sequence may be stretched or compressed (hence, ``warped'') in the temporal domain to match the other under certain constraints \cite[e.g.,][]{Muller2007}. The technique was first introduced by \cite{Bellman.Kalaba.1959} in adaptive control processes and later found several applications notably in speech and pattern recognition \cite[e.g.,][]{Sakoe1978, Myers1980}. More recently, DTW has been applied in space weather in particular for comparing modeled geomagnetic indices and solar wind data to observations \cite[e.g.,][]{Laperre2020, Owens2021, Bunting2022, Nilsson2022, Samara2022}. We first introduce its formulation and then describe the methodology for our application. 

The DTW technique finds the warping path between two sequences, e.g., $X = \{ x_1, x_2, ..., x_{|X|} \}$ and $Y = \{ y_1, y_2, ..., y_{|Y|} \}$, through the DTW cost matrix by choosing the path that minimizes the total cumulative cost compared to all other paths. The DTW cost matrix is defined as:
\begin{equation}
D(i,j) = \delta (i, j) + min \{ D(i-1, j-1), D(i-1, j), D(i,j-1) \} \label{eq:dtwcost}
\end{equation}
where $i, j$ are the indices of $X$ and $Y$, respectively, and $\delta (i,j) = |x_i - y_j|$ is the Euclidean distance between the element $x_i$ of series $X$ and the element $y_j$ of series $Y$. Here, the optimal warping path (i.e., alignment) can be found via back-tracing in the DTW cost matrix; this process is carried out by choosing the previous points with the minimum cumulative distance \citep[e.g.,][]{Berndt1994, keogh_derivative_2001, Gorecki:2013wi}. A main disadvantage of DTW is that an element $x_i$ may be mapped to several elements of $Y$ --- this problem is called ``pathological mapping'' or ``singularities'' \citep[e.g.,][]{Sakoe1978}. To alleviate this issue, certain constraints have been added such as the so-called windowing, slope weighting, and step patterns \citep[see][and references therein]{keogh_derivative_2001}. Various variations of DTW technique have also been developed to optimize the alignments between the sequence elements/points \citep[e.g.][and references therein]{keogh_derivative_2001, doi:10.1137/1.9781611972726.12, Keogh:2005aa, Salvador2007, Efrat:2007aa, furtunua2008dynamic, zhu2012novel, yadav2018dynamic}. 

Choosing an appropriate DTW technique is crucial as different algorithms can lead to different results. Here, we choose the FastDTW algorithm developed by \citet{Salvador2007} due to its multi-level approach. First, the time series are sampled down to a very coarse resolution. A warp path is found in this coarse resolution and projected onto a higher resolution. The warp path is then refined and projected again to a higher resolution. The process of projecting and refining is repeated until a warp path is found for the original resolution time series. Most importantly, since the technique initially finds a warp path in the coarse resolution, it is particularly appropriate to capture large-scale features in the time series, notably ``CIRs'' with large speed gradients. 

To compute a metric similar to the skill score using the (classic) DTW technique, \citet{Samara2022} defined the Sequence Similarity Factor (SSF) to quantify the similarity between the modeled and observed time series as: 
\begin{eqnarray}
SSF = \frac{DTW_{score}(O,M)}{DTW_{score} ({O, \langle O \rangle})}, 
\end{eqnarray}
where $DTW_{score}(O,M)$ is the DTW cost calculated between the observed ($O$) and modeled ($M$) time series, and the $DTW_{score} ({O, \langle O \rangle})$ is the DTW cost between the observed series and the baseline. Here, the baseline is given as the average of the observed time series ($\langle O \rangle$). The SSF is zero when the model forecast is perfect; it equals to unity when the forecast performs exactly the same as the baseline; and it is higher than unity when the forecast is worse than the baseline. 

The DTW cost (equation~\ref{eq:dtwcost}) calculated between two sequences is the sum of the length (i.e., the Euclidean distance) between all the optimal alignment pairs. Therefore, the existence of singularities, where a data point $x_i$ has several possible optimal alignments to $y_j$, increases the DTW cost. The baseline is an average of the observed time series --- it is a constant that is not varied with time for a considered interval. The DTW mapping between a model with time variation and a constant sequence (i.e., a straight line) would have several singularities. As a result, we cannot directly compare the DTW cost between two sequences with time variation to the DTW cost between a time-varying sequence and the (constant) baseline.  For this reason, we propose two new metrics as follows. First, a normalized cost or normalized distance can be defined as the ratio of the DTW cost to the number of all alignment pairs. This quantity has the unit of the quantity under consideration (similar to RMSE and MAE). Its value corresponds to the average distance of all the mapped alignments that encodes both temporal and spatial differences between the two sequences; we call this quantity ``DTW normalized distance''. The normalized SSF is then defined as: 
\begin{equation}
{SSF}_{normalized} = \frac{{DTW}_{score}(O,M) / {DTW}_{length} (O,M)}{{DTW}_{score} ({O, \langle O \rangle}) / {DTW}_{length} ({O, \langle O \rangle})}, \label{eq:nssf}
\end{equation}
where ${DTW}_{length}$ represents the number of all the alignments between two sequences. When the ${SSF}_{normalized}$ is smaller than unity, the forecast model performs better than the baseline. When it is less than unity, the forecast model performs worse than the baseline. We consider these two newly defined metrics besides the classic metrics in Section~\ref{subsubsec:longtermperm}. 


\section{Results} \label{subsubsec:int-results}

\subsection{Performance of Helio1D using the long-term data} \label{subsubsec:longtermperm}

\begin{table}
\caption{List of the Helio1D long-series intervals and the point-by-point metrics calculated using the Helio1D series in comparison with OMNI data for the same intervals.}             
\label{table:performance}      
\centering                          
\begin{tabular}{c c c c c c c c}        
\hline\hline                 
Interval  &	\multicolumn{1}{p{1.5cm}}{RMSE (km/s)} & \multicolumn{1}{p{1.5cm}}{MAE (km/s)} & \multicolumn{1}{p{1.5cm}}{FastDTW distance (km/s)} & Pcc & Skill score & \multicolumn{1}{p{1.5cm}}{$SSF_{normalized}$} \\    
\hline                        
Jan -- June 2005 & 175 & 145 & 89 & 0.05 & -0.79 & \textbf{0.77} \\
July -- Dec 2005	 & 179 & 145 & 105 & -0.07 & -1.22 & 1.03 \\
Jan -- June 2006 & 124 & 98 & 62 & 0.17 &-1.00 & \textbf{0.84} \\
July -- Dec 2006	 & 178 & 151 & 81 & -0.25	 &-1.36 & \textbf{0.80} \\
Jan -- June 2007 & 155 & 125 & 69 & 0.03	 &-1.11 & \textbf{0.77} \\
July -- Dec 2007	 & 112 & 94 & 71 & 0.43 &	-0.36	 & \textbf{0.94} \\
Jan -- June 2008 & 133 & 108 & 72 & -0.02 & -2.24 & 1.16 \\
July -- Dec 2008	 & 124 & 105 & 60 & -0.03	 & -2.75 & 1.22 \\
Jan -- June 2009 & 91 & 70 & 55 & 0.10 & -0.78 & 1.08 \\
July -- Dec 2009	 & 75  & 59 & 42 & 0.12 & -1.15	 & 1.01 \\
Jan -- June 2010 & 103 & 80 & 67 & 0.32 & -0.27 & \textbf{0.87} \\
July -- Dec 2010	 & 142 & 122 & 108 & -0.07 & -1.17 & 1.30 \\
Jan -- June 2011 & 159 & 128 & 97 & 0.07 & -0.38 & \textbf{0.83} \\
July -- Dec 2011	& 156 & 129 & 90 & 0.07 & -0.32 & \textbf{0.79} \\
Jan -- June 2012 & 167 & 134 & 101 & 0.08 & -0.30 & \textbf{0.80} \\
July -- Dec 2012	 & 140 & 110 & 79 & 0.16 & -0.19 & \textbf{0.84} \\
Jan -- June 2013 & 184 & 150 & 95 & 0.22 & -0.10 & \textbf{0.66} \\
\hline
July -- Dec 2017	 & 145 & 117 & 79 & 0.03 & -1.02 & \textbf{0.95} \\
Jan -- June 2018 & 123 & 97 & 58 & 0.01 & -0.82 & \textbf{0.79} \\
July -- Nov 2018	 & 102 & 80 & 59 & 0.17 & -0.74 & \textbf{0.90} \\
\hline                                   
\end{tabular}
\end{table}

\begin{figure}
\centering
\includegraphics[width=\columnwidth]{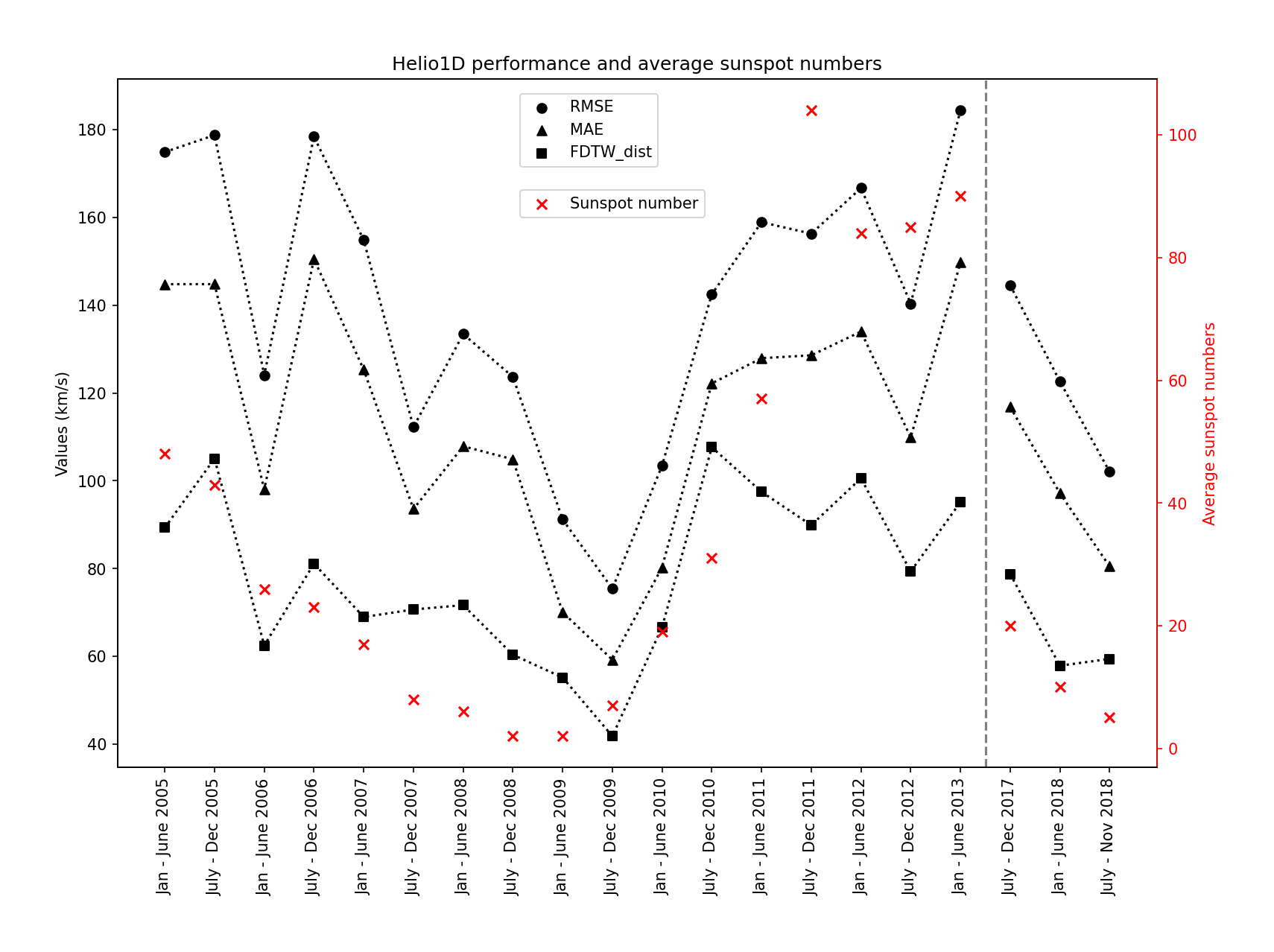}
\caption{The average RMSE (dot), MAE (triangle), and FastDTW normalized distance (square) versus the average sunspot numbers (cross) for eace 6-month interval. The values of the three metrics are shown on the left Y-axis (black) and the average sunspot numbers are shown on the right Y-axis (red). The data for 2005 - 2013 (solar cycle 23) and 2017 - 2018 (solar cycle 24) are separated by a vertical grey dashed line. The individual metric values for each solar cycle are connected by dotted lines.} 
\label{fig:performance-solar-cycle}
\end{figure}

\begin{figure}
\centering
\includegraphics[width=\columnwidth]{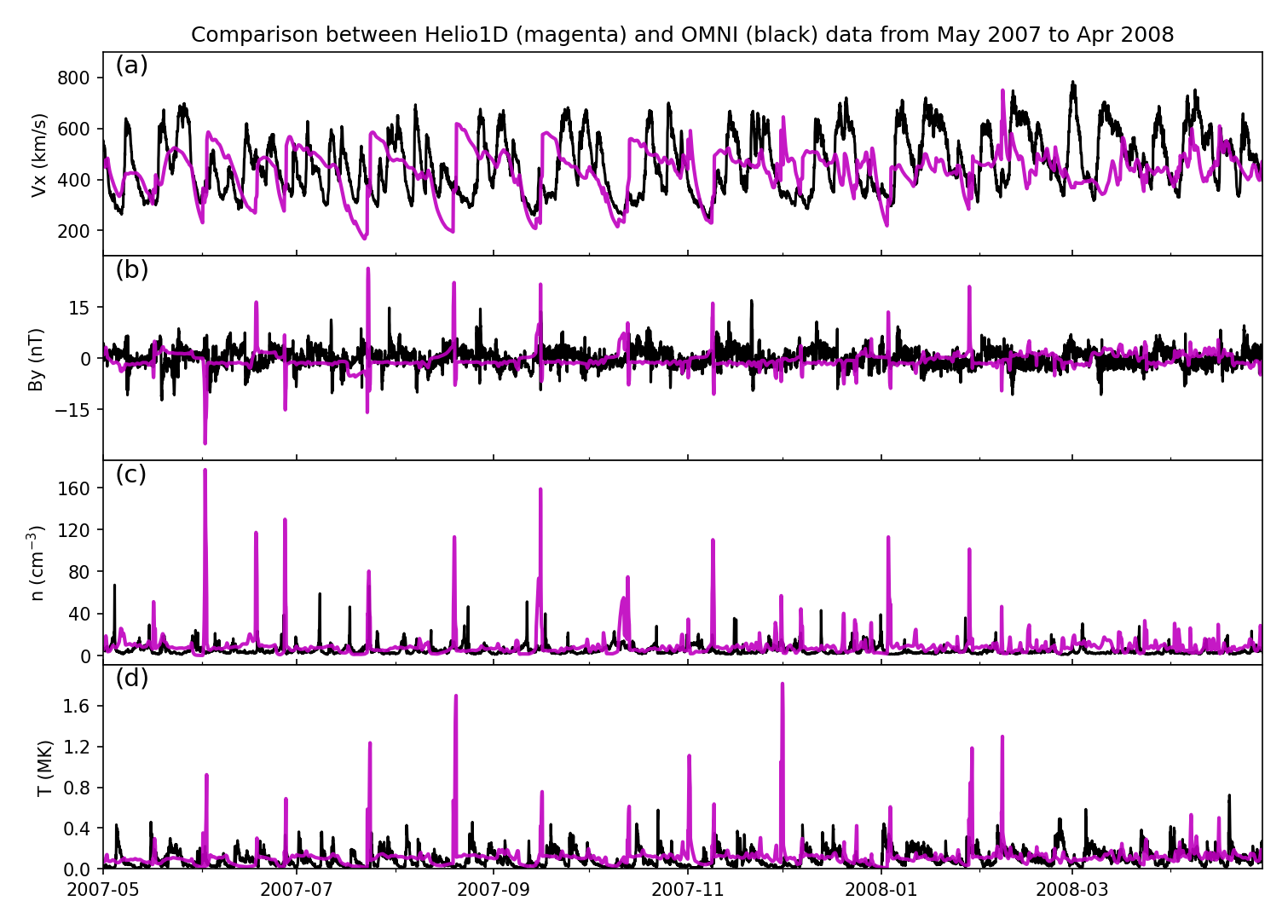}
\caption{Comparison of CIR modeling from the 1D MHD code (magenta) to the observations (black) from the OMNI database during May 2007 and April 2008 --- (a) radial bulk flow velocity ($V_x$); (b) tangential magnetic field component ($B_y$); (c) ion number density ($n$); and (d) ion temperature ($T$).} 
\label{fig:Helio1D-2007-2008}
\end{figure}

Table~\ref{table:performance} shows the RMSE, MAE, FastDTW normalized distance, Pcc, Skill score, and normalized SSF ($SSF_{normalized}$) from the comparison between the Helio1D and the in situ observations at L1 for each 6-month interval. The OMNI data were obtained at a 5-min resolution, and they are resampled to a 1-hour resolution to match the cadence of the Helio1D data. We find that the intervals in 2009 and the first half of 2010 have relatively low RMSE, MAE, and FastDTW distance, with the interval of July -- Dec 2009 having the lowest values for all three metrics. Overall, the FastDTW distance has a lower value than RMSE and MAE for each interval. The Pcc values are mostly close to zero, and reach negative values for certain intervals, indicating that there is no linear correlation between Helio1D and OMNI data (i.e., the two data are qualitatively dissimilar). The interval of July -- Dec 2007 has the best Pcc value of 0.43, showing somewhat linear correlation between the Helio1D model and the OMNI data; this interval is shown in Figure~\ref{fig:Helio1D-2007-2008}. The skill score is negative for all intervals, indicating that the Helio1D performs worse than the baseline model despite the relatively low errors or even positive Pcc, for some intervals. In contrast, the normalized SSF of most intervals has a value of less than unity (as highlighted in bold), indicating that the Helio1D indeed performs better than the baseline model. We discuss these different measures in Section~\ref{subsubsec:discussions}. To investigate the Helio1D performance in various phases of the solar cycle, we plot these metrics against the average number of the sunspots in Figure~\ref{fig:performance-solar-cycle}.

Figure~\ref{fig:performance-solar-cycle} shows the performance of the Helio1D characterized by the RMSE, MAE, and FastDTW distance for the intervals shown in Table~\ref{table:performance} (left y-axis, black) compared to the average sunspot numbers of the corresponding intervals (right y-axis, red). The sunspot number data were obtained as the monthly mean total sunspot number from the Royal Observatory of Belgium\footnote{Source: WDC-SILSO, Royal Observatory of Belgium, Brussels (https://www.sidc.be/silso/datafiles)}. The data in 2005 -- 2013 correspond to the solar cycle 23 and the data in 2017 -- 2018 correspond to the solar cycle 24. Here, we have more complete data for solar cycle 23 than the solar cycle 24. Based on the average sunspot numbers, we infer that our data for the solar cycle 23 comprise the declining phase from 2005 to 2007, the solar minimum from 2008 to mid-2009, the ascending phase from mid-2009 to 2011, and the solar maximum from 2012 to 2013. The rest of our data from mid-2017 to late 2018 correspond to the declining phase of the solar cycle 24. 

We find that all the metric values (shown in black) vary with the average sunspot numbers (shown in red). During the declining phase of the solar cycle 23, all the metric values proportionally decrease with the average sunspot numbers. During the ascending phase of the solar cycle 23, all the metric values roughly increase with the average sunspot numbers. The most striking features are during the solar minimum and the solar maximum. The metric values reach lowest values during the solar minimum and/or the late declining phase and early ascending phase. Here we find the minimum RMSE, MAE, and FastDTW distance six months after the lowest average sunspot number in Jan -- June 2009. The highest errors are found during the solar maximum, which are not surprising as the solar magnetic fields often undergo changes and they are thus less predictable. During the declining phase of the solar cycle 24, we find that the errors proportionally decrease following the average sunspot numbers. These results demonstrate that the performance of the Helio1D varies with the solar cycle --- it performs best during the solar minimum and it performs worst during the solar maximum. This finding is consistent with the performance of solar wind modeling by other existing models \citep[e.g.][]{2019SoPh..294..170H}. We further discuss these results in Section~\ref{subsubsec:discussions}.

To understand how the Helio1D pipeline qualitatively performs, we next visualize an example of the model output against the observations at L1. Figure~\ref{fig:Helio1D-2007-2008} shows the Helio1D output in magenta and the OMNI data in black. In Fig~\ref{fig:Helio1D-2007-2008}a, the bulk flow velocity displays a recurring pattern of the stream interaction regions (i.e., CIRs) characterized by the transition from slow to fast wind, clearly noticeable between August and November 2007. The stream interfaces also collocate with the polarity change in $B_y$, shown in Fig~\ref{fig:Helio1D-2007-2008}b, and the enhancement of density and temperature in Fig~\ref{fig:Helio1D-2007-2008}c and Fig~\ref{fig:Helio1D-2007-2008}d, respectively. Comparing to the observations, we find that the modeled solar wind speed agrees qualitatively well and that the model correctly produces consistent CIR formation. Nevertheless, the number density and temperature in Fig~\ref{fig:Helio1D-2007-2008}c and Fig~\ref{fig:Helio1D-2007-2008}d at the stream interfaces are generally higher than the observed values. These higher peaks are produced as a consequence of over-compression at the stream interfaces due to the ideal MHD plasma assumption in the 1D MHD code, which limits dissipation. Despite some over-compression and temporal uncertainties, we conclude that the interfacing between Multi-VP and 1D MHD works reasonably well. We further perform qualitative assessments of the pipeline performance and discuss calibration in Section~\ref{subsection:calibration}. 

From the long-term results (not shown), we find that the Helio1D pipeline correctly produces the expected large-scale modulation of CIRs especially in the bulk flow velocity, demonstrating that the Helio1D pipeline correctly produces long-term solar wind fluctuations. On a shorter timescale, there are some mismatches between the amplitudes of the fluctuations of the Helio1D results and the observations. We quantify these differences in detail in Section~\ref{subsec:perf-fdtw}. 

\subsection{Performance of Helio1D with the FastDTW technique}\label{subsec:perf-fdtw}

\begin{figure}[ht]
   \centering
   \includegraphics[width=\columnwidth]{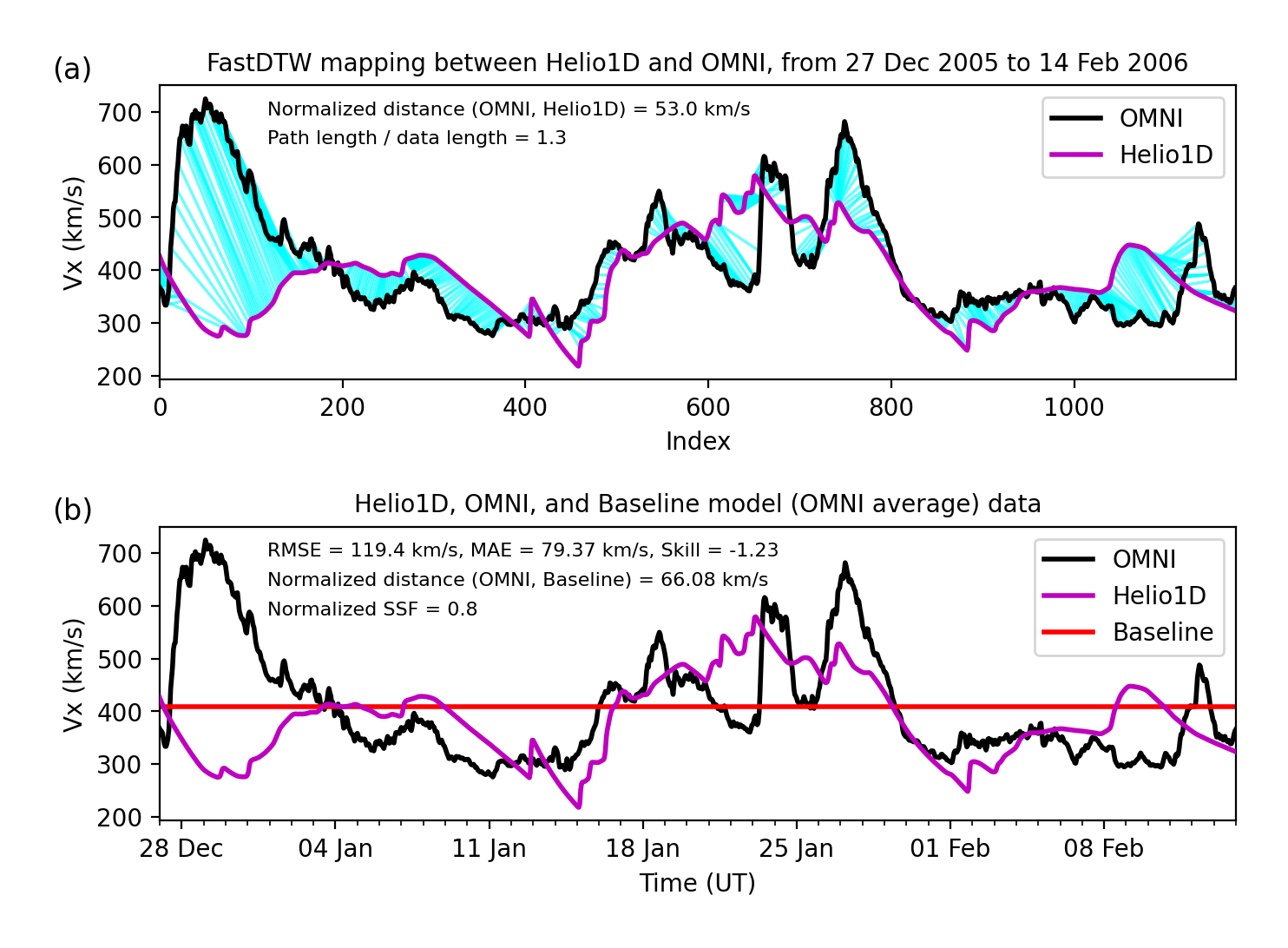}
   \caption{Example of the FastDTW mapping and performance metrics. (a) The FastDTW mapping (cyan) between the predicted CIR structures with Helio1D (purple) and the OMNI (black) data. (b) Helio1D, OMNI, and the baseline model (red) of the same interval. Relevant numbers indicative of the model performance are noted.} 
   \label{fig:DTW}
\end{figure}

Since the classic metrics only indicate crude information of the model performance when comparing to real data, we consider applying FastDTW algorithm. Fig~\ref{fig:DTW} shows an example of the FastDTW mapping between Helio1D and OMNI time series for the stream interfaces during 27 December 2005 and 14 February 2006. The OMNI data (black) show two clear stream interfaces and two corresponding high-speed streams. It can be seen that the Helio1D correctly produces the second stream interface along with the adjacent high-speed stream period compared to the observation despite some time delay and differences in the finer-scale structures.

Fig~\ref{fig:DTW}a shows the FastDTW alignments (cyan) that map large-scale features consisting in the stream interfaces and high-speed stream. The normalized FastDTW distance is about $53$ km/s for this interval. As there can be several alignments for a data point (e.g., at local peaks), the number of alignment pairs is equal or larger than the length of the data. We indicate the number of the alignment pairs (called ``path length'') to the length of the data in Fig~\ref{fig:DTW}a. Fig~\ref{fig:DTW}b shows the Helio1D, OMNI, and baseline model (red) of the same interval. We note the RMSE of $119.4$ km/s and MAE of $79.4$ km/s, measured between the Helio1D and OMNI data. The skill score, obtained from equation~\ref{eq:skill}, is $-1.23$ indicating that the Helio1D performs worse than the baseline model. Nevertheless, the normalized SSF (equation~\ref{eq:nssf}) of $0.8$, obtained from the comparison of the FastDTW mapping between OMNI and Helio1D to the FastDTW mapping between OMNI and baseline model, indicates that the Helio1D model performs better than the baseline model. 

This example shows that the Helio1D model prediction is still useful despite the negative skill score. We propose that the FastDTW algorithm can be used for mapping large-scale features, in particular, of the stream interfaces and high-speed streams. Here, the alignments can be used for extracting detailed information (i.e., the normalized FastDTW distance and normalized SSF) indicative of the model performance. We further exploit the alignments for model calibration in Section~\ref{subsection:calibration}.

As illustrated in Fig~\ref{fig:DTW}a, the FastDTW alignments optimally match large-scale features in the time series. For a given FastDTW map path, we may define the time difference as $\Delta t = t_{OMNI} - t_{1D}$ and the velocity difference as $\Delta V = V_{OMNI} - V_{1D}$, where the subscript OMNI means the observations and the subscript 1D means the Helio1D results. Here, the statistical information on $\Delta t$ and $\Delta V$ from the FastDTW mapping are useful for investigation of the model performance in terms of timing and magnitude difference. For each mapping path, $\Delta t > 0$, i.e., $t_{OMNI} > t_{1D}$, means that the Helio1D model result is ahead of the observation; while $\Delta t < 0$ means that the Helio1D model result is behind (e.g., lagged or delayed) with respect to the observation. Similarly, $\Delta V > 0$, i.e., $V_{OMNI} > V_{1D}$, means that the Helio1D model result underestimates the observed speed; while $\Delta V < 0$, i.e.,$V_{1D} > V_{OMNI}$, means that the Helio1D model result overestimates the observed speed. We can measure the model performance by constructing histograms of distribution of $\Delta t$ and $\Delta V$ as follows.

\begin{figure}[ht]
   \centering
   \includegraphics[width=\columnwidth]{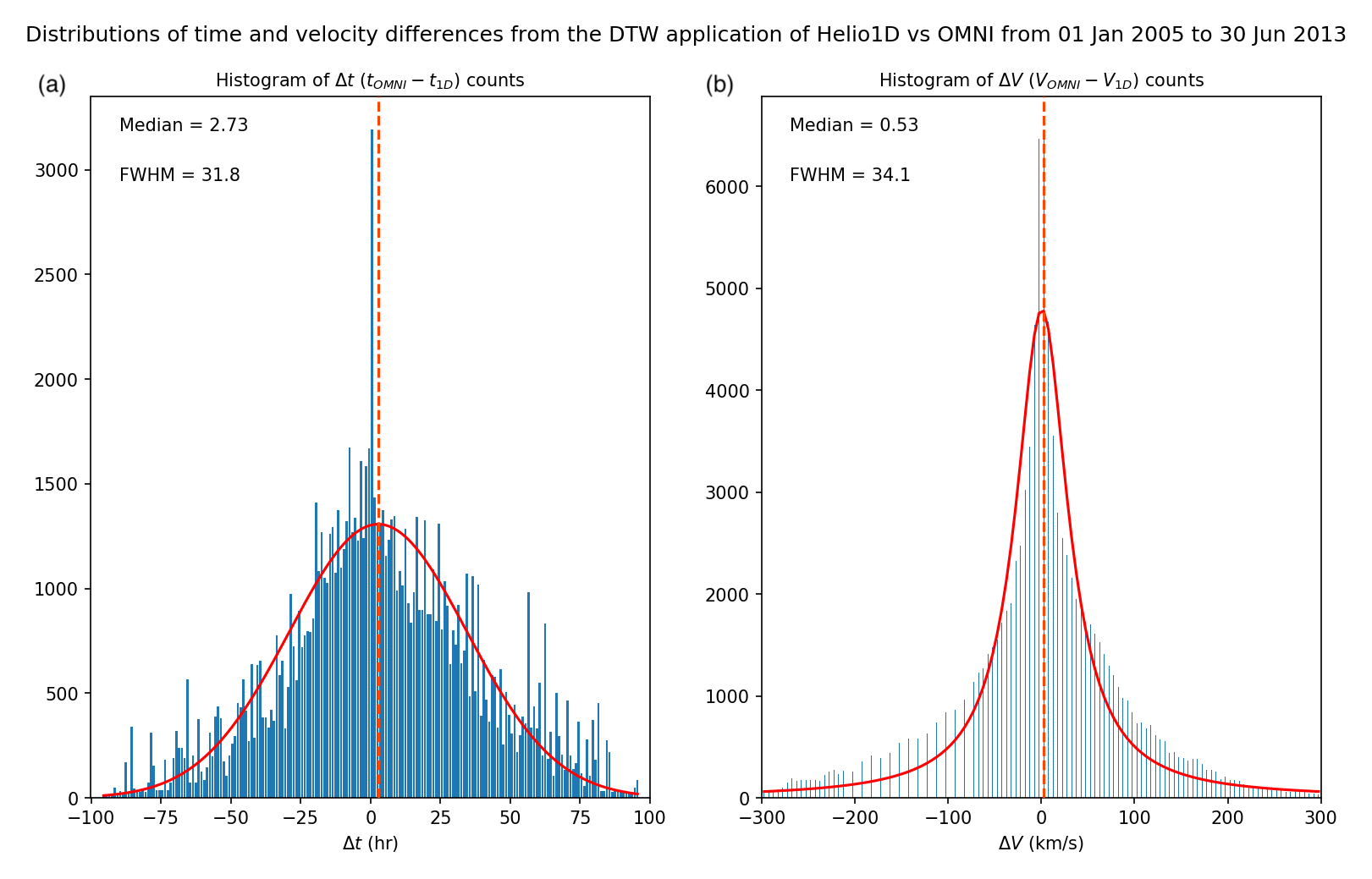}
   \caption{Histograms of (a) time delay and (b) velocity difference obtained from FastDTW alignments between the Helio1D and OMNI data from 2005 to 2018. The Gaussian (a) and the Cauchy-Lorentzian (b) distribution fits (red solid lines) are applied to obtain median (red vertical dashed line) and FWHM of the time delay and the velocity difference, respectively.} 
   \label{fig:hist-FDTW-fits-cycle23}
\end{figure}

Fig~\ref{fig:hist-FDTW-fits-cycle23} shows histograms of the $\Delta t$ and $\Delta V$ of all the data obtained from FastDTW alignments on the velocity time series from 2005 to 2013. Using the Gaussian distribution fit on the histogram of $\Delta t$ in Fig~\ref{fig:hist-FDTW-fits-cycle23}a, the median of the $\Delta t$ is about $2.7$ hours; the full width at half maximum (FWHM), i.e., $~ 2.355 \sigma$, where $\sigma$ is the standard deviation, is about $31.8$ hours.  In Fig~\ref{fig:hist-FDTW-fits-cycle23}b, we apply the Cauchy-Lorentzian to the $\Delta V$ distribution to extract their statistical properties as it best fits with the shape of the distribution. We find that the $\Delta V$ has the median of $0.53$ km/s and the FWHM of $34.1$ km/s. This means that there is almost no systematic bias for the modeling of the velocity amplitude by the Helio1D, with the majority of the modeling having $\Delta V$ within $\pm 34.1$ km/s. In brief, considering several years of statistics covering parts of the solar cycle 23, the Helio1D model results show minimal time and velocity differences compared to the observations.

\begin{figure}[ht]
\centering
\includegraphics[width=\columnwidth]{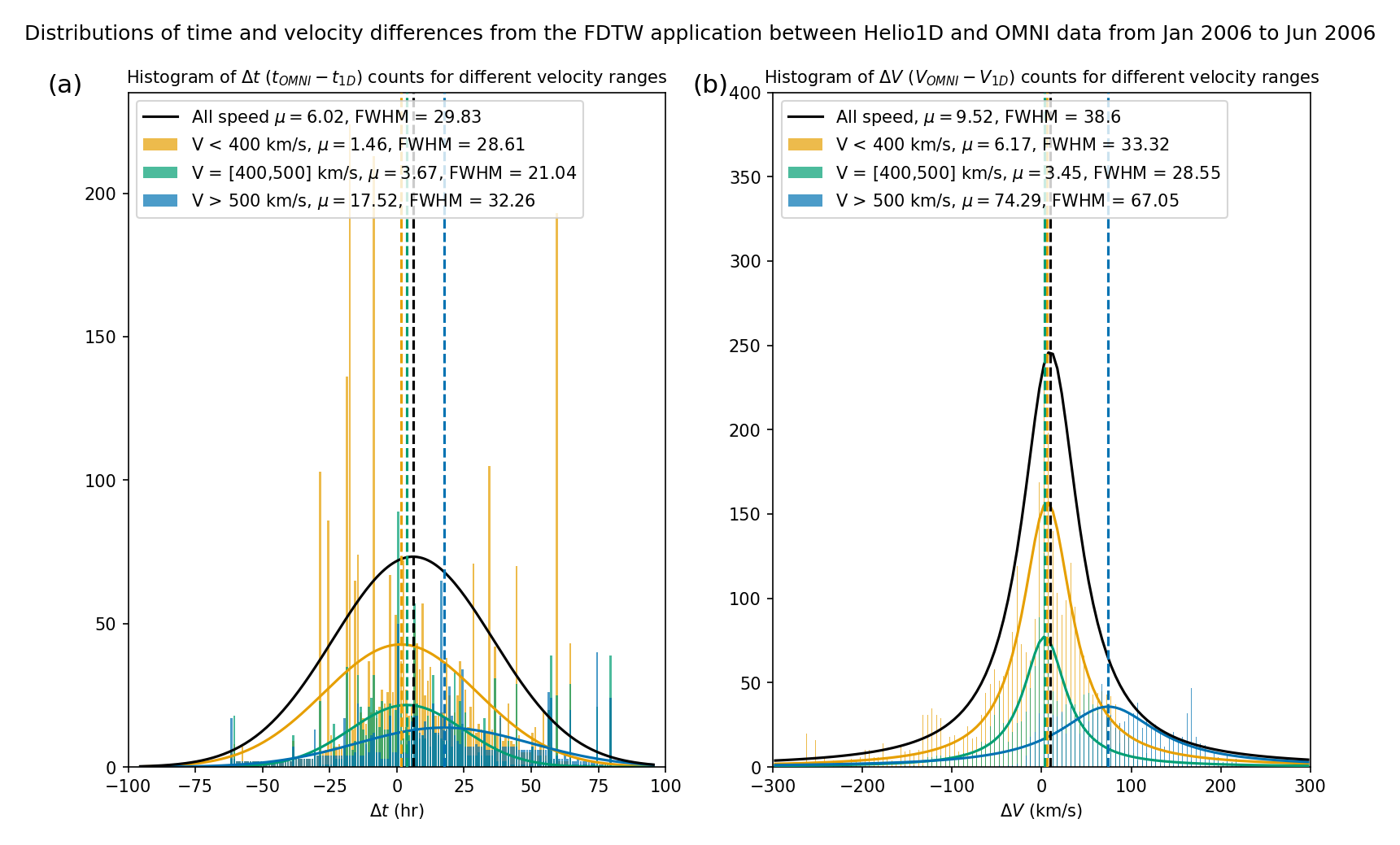}
\caption{Histograms of (a) time delay and (b) velocity difference obtained from FastDTW alignments between the Helio1D and OMNI data from Jan to June 2006. Blue, red, and green bars represent slow, intermediate, and fast winds, respectively. (a) The Gaussian distribution fits are applied to obtain median ($\mu$) and FWHM of the time delay. (b) The Cauchy-Lorentzian distribution fits are applied to obtain $\mu$ and FWHM of the velocity difference. The median and FWHM for all solar wind speeds are also given. See appendix for $\mu$ and FWHM all 6-month intervals.} 
\label{fig:hist-FDTW-fits}
\end{figure}

Using the FastDTW alignment information, we also explore the model performance for different types of the solar wind. In Fig~\ref{fig:hist-FDTW-fits-cycle23}b, for instance, we find that the $\Delta V$ distribution has slight skew or an asymmetry towards the negative values. This indicates that the Helio1D model likely overestimates the solar wind speed. To understand the model performance in detail, we consider histograms of $\Delta t$ and $\Delta V$ for the individual 6-month intervals in Table~\ref{table:performance}. Particularly, we categorize the solar wind schemes into three types: (1) slow wind with $V < 400$ km/s, (2) moderate wind with $400 \leq V \leq 500$ km/s, and (3) fast wind with $V > 500$ km/s. This categorization is based on the observed solar wind. Fig~\ref{fig:hist-FDTW-fits} shows an example of distributions of $\Delta t$ (a) and $\Delta V$ (b) with individual distribution fit functions to the slow (yellow), moderate (green), and fast (blue) solar winds, in addition to all solar wind (black). Table~1 in Appendix summarizes the statistical information of all the 6-month intervals. 

We now discuss our fitting results to the histograms of $\Delta t$ and $\Delta V$ obtained from the FastDTW algorithm in Fig~\ref{fig:hist-FDTW-fits} for the data in Jan - June, 2006. For the slow and moderate winds, the median time differences are within a few hours, with $\Delta t_{slow} = -1.5$ hours and $\Delta t_{moderate} = 3.8$ hours. For the fast wind, in contrast, we obtain a rather large time delay with $\Delta t_{fast} = 17.5$ hours. Overall, the FWHM of time delay is about $30$ hours for all solar wind speed ranges. For the velocity difference, we find that the slow and moderate winds have small medians of $\Delta V$, with $\Delta V_{slow}  = 6.2$ km/s and $\Delta V_{moderate}  = 3.5$ km/s. For the fast wind, however, the median of $\Delta V$ is bigger, with $\Delta V_{fast}  = 74$ km/s. This example shows that the Helio1D indeed performs differently for various solar wind speed ranges. For this interval, for instance, the Helio1D model underestimates the speed of the fast wind. To investigate whether this particular behavior persists, we next consider the median $\Delta t$ and $\Delta V$ of the individual 6-month intervals throughout the solar cycle 23. 

\begin{figure}[ht]
\centering
\includegraphics[width=\columnwidth]{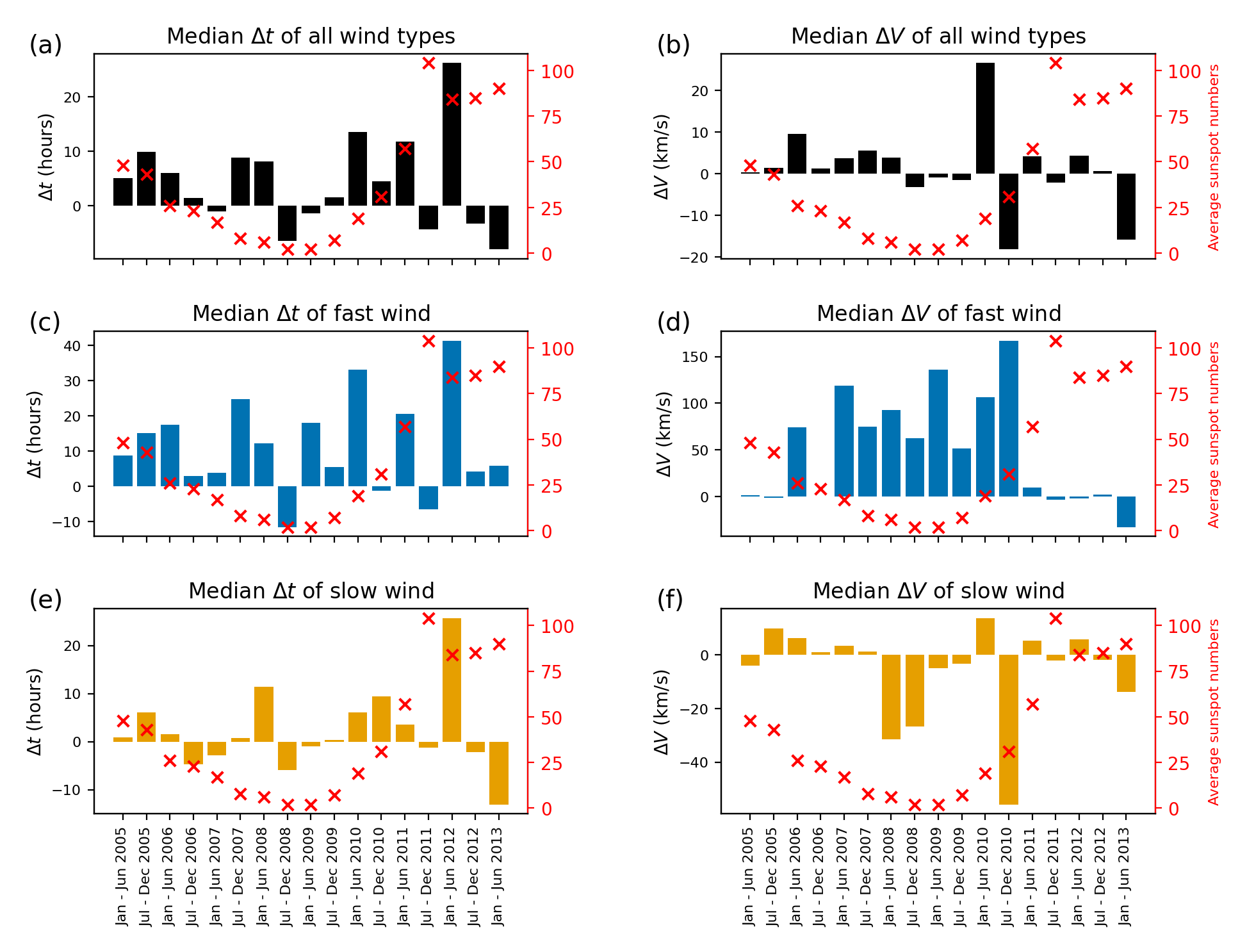}
\caption{Bar plots of the median time difference ($\Delta t$; left panels) and velocity difference ($\Delta V$; right panels) from the FastDTW technique applied to the 6-month intervals in 2005 - 2013, over-plotted by the average sunspot numbers (red cross, right Y-axis) indicative of the solar cycle. The bar plots are shown for (a, b) all wind types, (c, d) slow wind, and (e, f) fast wind.  } 
\label{fig:bar-FDTW-cycle23}
\end{figure}

Fig~\ref{fig:bar-FDTW-cycle23} shows bar plots of the median $\Delta t$ in the left panels and $\Delta V$ in the right panels, separated for all wind types (a, b), fast wind (c, d), and slow wind (e, f). The bar plots are shown against the average sunspot numbers of the corresponding intervals to indicate the phases of the solar cycle. We note that the moderate wind does not show a particular trend and thus excluded from this plot. Overall, we find that the Helio1D has a positive $\Delta t$ (Fig~\ref{fig:bar-FDTW-cycle23}a), indicating that the Helio1D timing is ahead of the observations. The largest positive $\Delta t$ is found during the ascending phase and the solar maximum (Jan - June 2012 in Fig~\ref{fig:bar-FDTW-cycle23}a). This positive $\Delta t$ is larger for the fast wind (40 hours) compared to the slow wind (25 hours) as seen in Figs~\ref{fig:bar-FDTW-cycle23}c and \ref{fig:bar-FDTW-cycle23}e. For the velocity difference for all wind types, we find that the Helio1D provides a minimal velocity difference within 10 km/s except during the ascending phase and solar maximum as seen in Fig~\ref{fig:bar-FDTW-cycle23}b. When considering the different wind types, we find that the median $\Delta V$ of the fast wind (Fig~\ref{fig:bar-FDTW-cycle23}d) is markedly different from that of the slow wind (Fig~\ref{fig:bar-FDTW-cycle23}f). In particular, during the solar minimum and early ascending phase, there is a large positive $\Delta V$, indicating that the Helio1D underestimates the speed of the fast wind ($V_{1D} < V_{OMNI}$) by 50 - 150 km/s. In contrast, during the same intervals, there is mostly negative $\Delta V$ for slow wind. In other words, the Helio1D overestimates the speed of the slow wind ($V_{1D} > V_{OMNI}$). This overestimation can go up to 60 km/s as seen during the ascending phase.

In brief, we find that the timing of the Helio1D results (i.e., stream interfaces) is often ahead of the observations (consistent with Fig.~\ref{fig:Helio1D-2007-2008}). This positive $\Delta t$ is even larger for the fast wind. In terms of the velocity, we find that the Helio1D model underestimates the fast wind speed while overestimates the slow wind speed. This effect is clear during the solar minimum and the early ascending phase.

\section{Pipeline calibration} \label{subsection:calibration}

\begin{figure}
\centering
\includegraphics[width=\columnwidth]{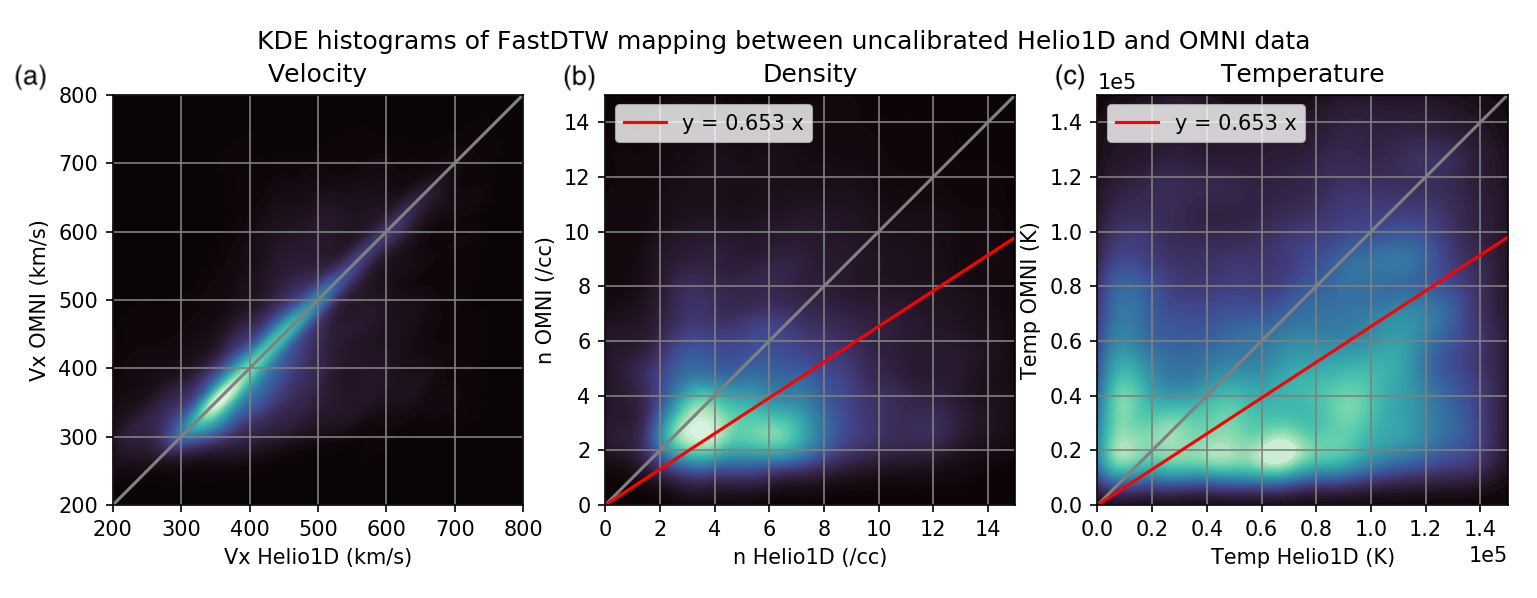}
\caption{Calibration plots obtained from the FDTW alignments between Helio1D (horizontal axis) and OMNI (vertical axis) time series for (a) solar wind speed, (b) number density, and (c) temperature.} 
\label{fig:DTW-hist2d}
\end{figure}  

As mentioned in Section~\ref{subsubsec:longtermperm}, Helio1D usually overestimates the number density and temperature of plasma at stream interaction regions as a consequence of over-compression at the stream interfaces. To alleviate this issue, we consider a post-calibration of the data to lower the extreme peaks in the Helio1D outputs. As shown in Section~\ref{subsec:perf-fdtw}, FastDTW provides the alignments between the modeled and observed solar wind speeds. Since the FastDTW maps similar structures within the two time series, we create a calibration plot for the mapped values between the Helio1D and OMNI data. Fig~\ref{fig:DTW-hist2d} shows a kernel density estimation (KDE) plot of the mapped solar wind speeds from the Helio1D (horizontal axis) and the OMNI data (vertical axis), which shows their smoothed 2D density probability histogram. The brighter color in the plot corresponds to a higher probability density. Here, the velocity data in Fig~\ref{fig:DTW-hist2d}a has a high probability density that mostly follows the line of slope of 1. This demonstrates that there is overall no bias from the Helio1D modeling (similar to Fig~\ref{fig:hist-FDTW-fits}b), and that the magnitude of the solar wind speed from Helio1D mostly matches the OMNI data.

The FastDTW alignments information allows us to map not only velocity but also plasma density and temperature values between the modeled and observed time series. In Figs~\ref{fig:DTW-hist2d}b and \ref{fig:DTW-hist2d}c, we compare the plasma number densities and temperatures between the Helio1D and OMNI using the mapped path obtained from the FDTW application on the solar wind speed. It is clearly visible that the number density and the temperature from Helio1D are higher than observed values. To lessen these overestimations, we consider using linear regression. Here, we apply a linear fit onto the data shown as the red solid line in Fig~\ref{fig:DTW-hist2d}b. Similarly, we apply a linear fit function to the plasma temperature in Fig~\ref{fig:DTW-hist2d}c. The linear fit functions of both number density and temperature are found to have  a slope of 0.653. Using these linear functions, we find that the magnitudes of the peaks of the number density and temperature better match those of the observations (not shown). We thus conclude that our calibration on the plasma number density and temperature with the linear function significantly improves the Helio1D modeling. We apply these calibrations for the number density and temperature of the Helio1D pipeline for operational forecasting described in Section~\ref{sec:operation}.


\section{Operational forecasting} \label{sec:operation}

We now shift our focus to the development of a stable ``prototype'' service for solar wind forecasting with Helio1D. To continuously predict solar wind conditions, several technical aspects as mentioned in Section~\ref{subsec:interfacing} are addressed. First and foremost, characterization of model uncertainties is critical to assess the modeling results in order to plan for mitigation of plausible impacts of extreme events. We provide model uncertainties using an ensemble modeling in Section~\ref{subsec:ensemble}. Secondly, the 1D MHD model requires that the length of input time series covers at least one solar rotation in order to produce consistent CIR formation. The Multi-VP model has a setup that provides for the daily solar wind emergence comprising 72 hours of solar wind conditions covering from the present day ($D$) to the next two days ($D+2$). Therefore, the solar wind emergence time series from Multi-VP must be concatenated to produce a continuous time series with a minimum length of 27.25 days. Lastly, there can be data gaps and/or unphysical values arising from either the inputs for MultiVP (i.e., magnetogram), due to the lack of data or the presence of invalid data points, or numerical errors. The implementation of automatic procedures to tackle these issues are detailed in Section~\ref{subsec:technical}.

\subsection{Helio1D ensemble modeling with 21 virtual targets}\label{subsec:ensemble}

We perform an ensemble forecasting of the MULTI-VP and 1-D MHD models by considering a range of heliospheric latitudinal and longitudinal uncertainties. For example, a solar wind structure that is ahead or behind in time compared to the observation can be accounted for by considering the time-series at some nearby heliospheric longitudes. Also, the magnitude of the solar wind speed profiles is different at different heliospheric latitudes as a function of the warping of the Heliospheric Current Sheet and nearby velocity gradients. Using daily magnetograms from WSO, we set up the Multi-VP model to automatically generate daily one-dimensional solar wind profiles that cover from $D$ to $D+2$ at the sub-Earth point and the surrounding virtual targets. The surrounding virtual targets are set to spread up to $15^o$ from the sub-Earth point in latitude and longitude; they are set at $5^o$ apart from each other in latitude and longitude; these chosen points constitute the star-grid pattern with the total of 21 points including sub-Earth point (see Appendix). The spatial cuts through these 21 points on the 2D map of solar wind emergence from Multi-VP along the ecliptic are then translated to temporal solar wind profiles. These 21 time-series inputs are automatically fed into the 1D MHD to propagate them in parallel from 0.14 AU to 1 AU. The 1D MHD model is rather computationally inexpensive; the model running with 21 virtual targets remains fast compared to 3D MHD modeling in general. 

\begin{figure}[ht]
\centering
\includegraphics[width=\columnwidth]{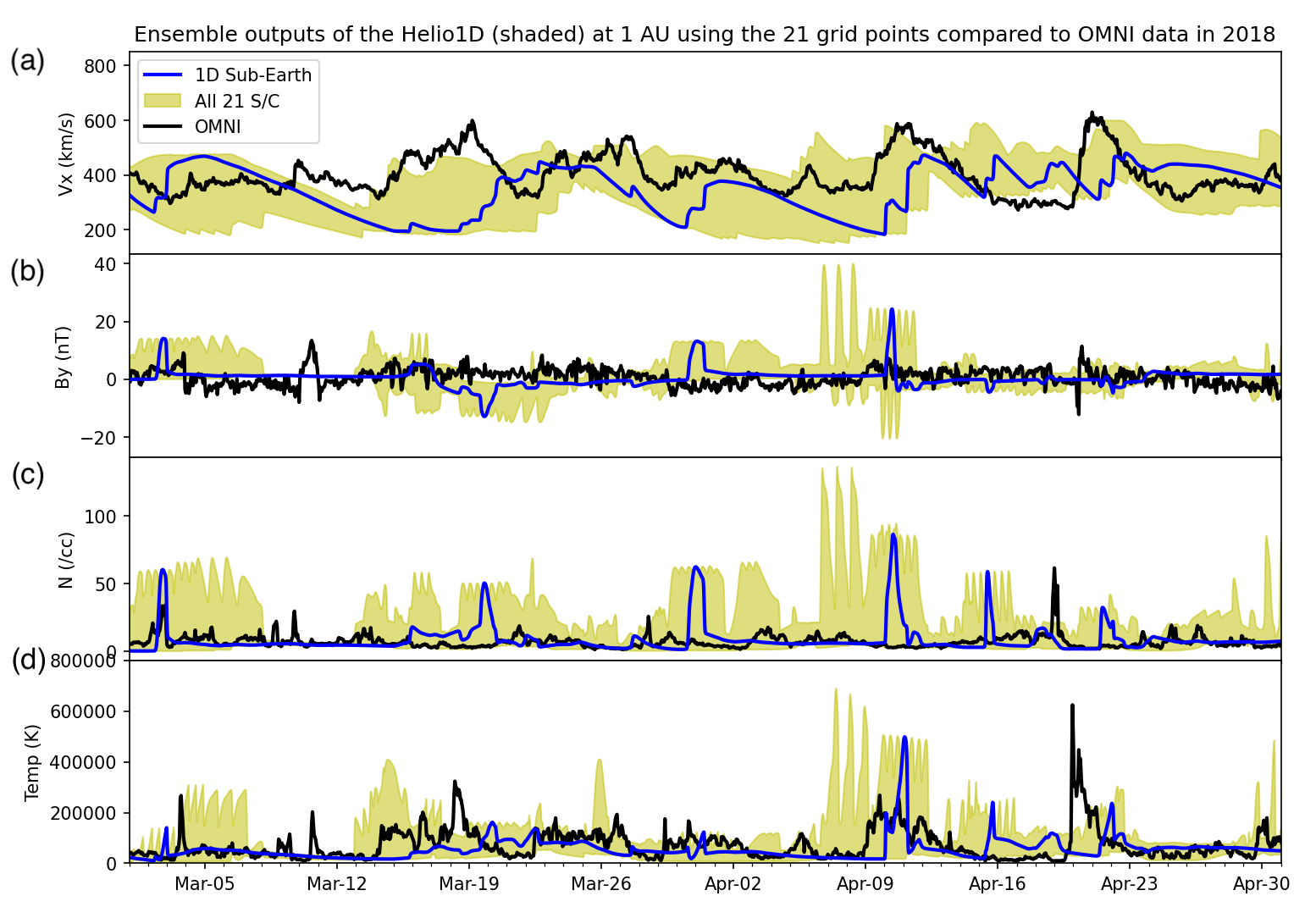}
\caption{Helio1D ensemble time series outputs using the 21 virtual spacecraft compared to the OMNI data for March to April, 2018. The ensemble Helio1D outputs are shaded in green and the output at sub-Earth point is plotted in blue. The observation data are represented by a black line.} 
\label{fig:Helio1D-ensemble}
\end{figure}

Fig~\ref{fig:Helio1D-ensemble} shows results from the Helio1D pipeline with the 21 virtual points between March and May, 2018. The Helio1D data at the sub-Earth point are shown in blue, and the observation data are shown in black. The green shade highlights the spread of values between the minimum and maximum among the 21 virtual targets at each hour. Panels (a) - (d) show the solar wind speed, tangential magnetic field, plasma number density, and plasma temperature, respectively. We note that the post-calibration with the linear functions for number density and temperature (found in Section~\ref{subsection:calibration}) has been applied. Here, the ensemble spread in green shade provides an error bar. In this example, we find that the stream interfaces modeled for the sub-Earth point appear to lag behind the observations for about two days. Nevertheless, the timing uncertainties from the values at the virtual points indeed cover the timing of the arrivals of observed stream interfaces. Overall, we find that most of the observed data points fall within the error spread for both magnitude and timing, although there are some underestimations of the speed of the high-speed stream (consistent with our findings in Section~\ref{subsec:perf-fdtw}). We conclude that the Helio1D modeling using the 21 virtual targets improves the modeling compared to using only one virtual point targeting at Earth.

\subsection{Solar wind data concatenation \& Data gaps} \label{subsec:technical}

The main technical problems deal with the automated interfacing between Multi-VP and 1D MHD model. Unlike the model benchmarking and case-by-case analysis, we rely on automatic modeling for both the Multi-VP and 1D MHD models. Here, the Helio1D pipeline is scheduled to run in-house every day. First, the Multi-VP is set up to produce a daily forecast with a data length of 3 days, covering from day $D$ to $D+2$. We first concatenate these Multi-VP data for each virtual target by averaging the data from the day $D-30$ to the present day $D$ to make the time series input to be sufficiently long. After the concatenation process, this 1-month solar wind profile at 0.14 AU is subsequently propagated by the 1D MHD model to provide the solar wind profile at 1 AU. Since the solar wind takes time to propagate to the Earth, we gain an extra lead time of 2 – 7 days depending on the solar wind speed. To provide a total lead time of 4-days from Helio1D, we limit the extra time gained from the 1D MHD propagation modeling to 2 days. This setup has been done to automatically interface the Multi-VP and 1D MHD to provide daily solar wind nowcast and forecast from day $D$ to $D+4$ (see branch (b) in Fig~\ref{fig:helio1d-pipeline}). In an absence of the magnetogram data, the magnetogram of the previous solar rotation is programmed to be fetched. This procedure is implemented based on the assumption that the coronal structures remain unchanged compared to the previous solar rotation (i.e., when the consistent, recurring CIRs are expected).

Furthermore, one of the models, or both, may terminate earlier than expected and give incomplete outputs. This is because, in reality, there can be data gaps arising from the lack of daily data (e.g., magnetogram data), or unphysical values in the data. The latter can come from numerical errors from one of the codes or poor quality of the raw inputs. To prevent the Helio1D pipeline from early termination, we perform an automatic search for data gaps and unphysical values (after performing the fetching of previous magnetogram when the present-day magnetogram is unavailable as mentioned above). The data gaps are then filled using linear interpolation between two available data points. The unphysical values identified using the criteria (see Appendix) are subsequently removed. The resulting data gaps are then filled using linear interpolation. To remove unphysical features that may arise from these procedures, we perform a running average on the data using a window of $\pm 3$ hours centered around the data point. Finally, when either the input or daily output is shorter than expected, we fill the data with latest available values to complete the length of expected output. For practical purpose, we provide daily solar wind time series extending from day $D-3$ to $D+4$ for a given day $D$.

Example plots of the daily data from the Helio1D are publicly available; they can be accessible from \url{http://swift.irap.omp.eu/} and, in the near future, through the ESA Space Safety Program via \url{www.esa.int/Space_Safety/Space_weather}.

\section{Discussion} \label{subsubsec:discussions}

We developed the Helio1D pipeline for solar wind modeling by interfacing Multi-VP for the coronal part and 1D MHD for the interplanetary part. The interfacing has been done for the first time as an effort to connect two developed models for operational forecasting purpose. Using the long-term Multi-VP data produced with the WSO magnetograms in 2005 - 2013 and 2017 - 2018, we evaluated the pipeline performance in various phases of the solar cycles 23 and 24. In addition to classic metrics such as RMSE, MAE, and skill score, we devised a new metric based on the FastDTW algorithm. Using the long-term data and the metrics, we discuss our main findings as follows.

\begin{itemize}
\item[1.] The performance of the Helio1D pipeline is dependent on the phases of the solar cycle. Using the RMSE, MAE, and FastDTW distance, measured for each 6-month interval, we find that their values positively correlate with the average sunspot numbers of the previous 6-month interval. The RMSE, for example, reaches a minimum value of 80 km/s during the solar minimum, and reaches a maximum value of 180 km/s during the solar maximum. Overall, the Helio1D pipeline provides minimal magnitude errors during the late declining phase and the solar minimum. 

\item [2.] The Helio1D pipeline produces consistent CIR formation. For each CIR, the solar wind speed profile at the stream interface qualitatively agrees with the observations (Fig~\ref{fig:Helio1D-2007-2008}a). Thus, the model sufficiently includes the large-scale physics of the CIRs. However, due to the limited dimensionality and ideal MHD assumption, the compression at stream interfaces (resulting in extreme peaks of number density and temperature) is rather strong compared to the observations. This effect was found to be common to 1D MHD models used for solar wind propagation \citep[e.g.,][]{Zieger2008}. To alleviate this effect, we perform a post-calibration to lower the peaks at stream interfaces using a linear function. This linear function was found using the FastDTW alignments of the solar wind profiles, which particularly mapped stream interfaces. 

\item[3.] Using the FastDTW alignments, we also evaluated the timing and magnitude differences between the Helio1D and the observation data. For the data in the solar cycle 23, we find minimal $\Delta t$ of 2.7 hours and $\Delta V$ of 0.5 km/s. However, when separating the data into shorter intervals of 6-months and categorizing the solar wind types based on its speed, we find different results summarized as follows. 
	\begin{itemize}
	\item[3.1] For all wind types, we find that the Helio1D timing is often ahead of that of the observation for about 10 hours. This timing error is highest six months after the average sunspot number reached the maximum in late 2011. Meanwhile, the $\Delta V$ for all wind types remains low throughout the solar cycle except for the ascending phase of the solar cycle in 2010. 
	\item[3.2] The fast wind has higher $\Delta t$ and $\Delta V$ than those of the slow wind. The Helio1D timing of fast wind is often 10 - 20 hours earlier than that of the observation, and this timing difference goes up to 40 hours six months after the highest sunspot number. For $\Delta V$, the Helio1D pipeline mostly underestimates the speed of the fast wind for about 50 - 100 km/s during the declining phase and solar minimum; this difference goes up to $\sim 150$ km/s during the late ascending phase. 
	\item[3.3] The timing of the slow wind agrees with that of the observations within $\pm 10$ hours, except during the solar maximum. For $\Delta V$, the Helio1D pipeline often overestimates the speed of the slow wind for about 20 - 40 km/s. The $\Delta V$ is largest with a value of $- 60$ km/s in late 2010.
	\end{itemize}
	
\end{itemize}

Comparing to existing models that have been benchmarked using large data sets, the performance of the Helio1D pipeline is in agreement with other equivalent models. \cite{2017SpWea..15.1461O} utilized solar wind emergence at 0.14 AU from MAS based on Carrington maps (e.g., WSO) of the photospheric magnetic field. Particularly, they used a large ensemble (N = 576) of solar wind time series from MAS, where the ensemble members are produced from sampling solar wind solutions within a range of latitudes about the sub-Earth point. The solar wind flows at 0.14 AU are then propagated using the HUX model \citep{2011SoPh..270..575R} that is based on the fluid momentum equation \citep[e.g.,][]{1978JGR....83.5563P} and takes into account a residual solar wind acceleration \citep{Schwenn1990}. Using a long interval of data from 1996 to 2016, they find that the median RMSE of the upwind ensemble propagation is 107 km/s. \cite{2020ApJ...891..165R} performed a forecasting of ambient solar wind using the WSA model and the HUX tool as well as the Tunable HUX \citep{2020ApJ...891..165R} to map solar wind flows from near Sun to Earth for the period 2006 - 2015. Their RMSE is ranging from 90 to 122 km/s while their MAE is ranging from 72 to 93 km/s. Their skill scores are negative for WSA/HUX (greater than -0.6) and positive (less than 0.06) for WSA/THUX. Compared to their studies, our Helio1D pipeline performance characterized by the average RMSE and MAE are in agreement with their models despite somewhat larger values by about a few 10 km/s. Nevertheless, RMSE and MAE are rather crude metrics; they do not tell all qualities of the prediction and the measurement, e.g., as discussed by \cite{https://doi.org/10.1029/2018SW002040} and several others. 

Regarding to the variation of the Helio1D performance with phases of the solar cycle, we find that our findings are generally consistent with other 1D MHD models. \cite{Zieger2008} performed an extensive validation of solar wind propagation using a 1D MHD model. They find that the variation of the coronal structure on the timescale of a solar rotation, characterized by the recurrence index of solar wind speed, plays an important role in the prediction efficiency. This explains our Helio1D results such that, in an absence of variation, i.e., during the late declining phase, the model generally predicts CIR formation consistent with the observations. In contrast, in a presence of strong variations, i.e., when there are CME emergences during the ascending phase and the solar maximum, the model generally predicts poor results compared to the observations. 

In terms of timing, we find that there is a bias on the solar wind stream arrival time such that the Helio1D solar wind streams usually arrive 10 hours earlier than the observations. This bias is stronger for the fast wind compared to the slow wind. This effect may come from the assumption of an ideal MHD plasma in the 1D MHD model that ignores finer-scale physics and/or interactions that could slow down fast wind or accelerate slow wind in the interplanetary space. Moreover, the limited dimensionality of the simulation does not take into account the propagation in other directions; this may result in the timing mismatch. In terms of magnitude, the Helio1D underestimates the speed of the fast wind while overestimates the speed of the slow wind. The bias on the fast and slow wind magnitudes may come from the lack of the physics on solar wind acceleration. In HUX model, an empirical, ad-hoc solar wind acceleration is incorporated into the model \citep{Schwenn1990}. Since the 1D MHD code was originally developed for solar wind propagation in the outer heliosphere \citep{2005JGRA..11011208T}, this physics was deemed less important and it was not added into the model. We note that this aspect must be addressed in a future improvement of the 1D MHD model for the inner heliosphere; this aspect out of scope of this work which focuses on interfacing the matured models.     

Our work devised a new exploitation of the FastDTW technique for a detailed assessment of the Helio1D performance. In particular, the FastDTW alignments encoded optimal time delays and magnitude differences between the modeled stream interfaces and observed structures. We highlight that the FastDTW can be used to provide more qualitative measures than RMSE, MAE, and skill score, for example. We also demonstrated that the FastDTW (normalized) distance correlates with the RMSE and MAE. Nevertheless, the FastDTW technique is not without caveats. The main constraint of DTW techniques in general is the pathological mapping or the singularities. Therefore, the obtained $\Delta t$ and $\Delta V$ from the DTW applications are not unique but instead depending on the constraints that have been added. In our case, we restrict that the alignments along the time axis must be within 96 hours (corresponding to the time that a solar wind with an intermediate speed should take to travel from 0.14 to 1 AU). Furthermore, we chose the FastDTW technique in particular because it searches to first map the structures at coarse scales, then refine to smaller scales. We emphasize that this algorithmic feature is desirable for mapping large-scale solar wind structures such as CIRs or CMEs.

Finally, we highlight the work that needs to be taken to develop an automated solar wind forecasting pipeline of Helio1D. In addition to the pipeline calibration to alleviate the pipeline caveats, there are other aspects that must be addressed. This comprises (1) providing model uncertainties and (2) dealing with data gaps and bad data. The Helio1D modeling of an ensemble of 21 solar wind solutions was shown to be a reasonable method to provide worst case scenario especially for the timing of stream interface arrivals. We note that the ensemble members can be further exploited to obtain an optimized forecast; we leave this aspect for future work. Most importantly, to develop a reliable service, we implemented strategies to tackle with data gaps and bad data points so that the pipeline can automatically run and provide continuous nowcast and forecast. This aspect should be further tested in order to evaluate their impacts on accuracy, performance, and stability of the operational forecasting pipeline. Such a task is of particular importance; it requires dedicated tests on this prototype pipeline in various situations before employing it for applications in real world. 

\section{Summary}

We developed a prototype pipeline ``Helio1D'' that automatically interfaces the coronal model ``Multi-VP'' and the solar wind propagation model ``1D MHD'' to provide nowcast and forecast of ambient solar wind containing CIRs at L1. The operational prototype pipeline provides daily solar wind modeling with a lead time of 4 days. In this work, we benchmarked and extensively tested the Helio1D pipeline using 10 years of data spanning from late 2004 - 2013 and 2017 - 2018. We evaluated the performance of the pipeline using classic metrics including RMSE, MAE, and skill score. In particular, we devised a new exploitation of the FastDTW technique to map large-scale solar wind structures of CIRs. We demonstrated that this technique can be used to characterize the detailed performance of the solar wind modeling with Helio1D. For instance, we characterized the statistical information on timing and magnitude differences for velocity profils of stream interfaces. Using this new approach, we assessed the pipeline performance for various phases of the solar cycle 23 and investigated the modeling bias for fast and slow solar winds. We find that the Helio1D pipeline performs best during the declining phase and solar minimum. However, the Helio1D pipeline often underestimates the speed of the fast wind while overestimates the speed of the slow wind, especially between the late declining phase and early ascending phase of the solar cycle 23. Furthermore, the solar wind structures modeled by Helio1D often arrives earlier than the observations especially for the fast wind. These caveats plausibly arise from the simplistic assumptions in the 1D MHD model, comprising the limited dimensionality, the lack of dissipation, and the solar wind acceleration physics, for example. Nevertheless, the Helio1D pipeline models consistent CIR formation while remains computationally light, which is desirable for operational forecasting. 

To transition from case-by-case benchmarking to operational solar wind forecasting, we implemented (1) the pipeline calibration to alleviate the over-compression at stream interface due to the ideal MHD assumption, (2) the ensemble modeling of 21 solar wind solutions at virtual targets including the sub-Earth point to provide timing and magnitude uncertainties, and (3) the procedures to remove unphysical data points and automatically fill data gaps. The latter two procedures were discussed to be crucial in providing continuous, reliable service while providing forecasting uncertainties. Alternatively, the Helio1D pipeline can be adapted to use other more reliable magnetogram sources and with higher resolution, e.g., from Air Force Data Assimilative Photospheric Flux Transport \citep[ADAPT;][]{2015SoPh..290.1105H, 2013AIPC.1539...11A, 2000SoPh..195..247W}. We emphasized that the implementation of procedures for making the pipeline fully operational, in addition to the pipeline benchmarking, are critical and they must be further tested in several situations before employing the pipeline for real applications.

\section*{Acknowledgements}
The SafeSpace project has received funding from the European Union’s Horizon 2020 research and innovation programme under grant agreement No 870437. The 1D MHD model of solar wind propagation initially developed by \citet{2005JGRA..11011208T} and used for Helio1D is duplicated from the Heliopropa service provided by CDPP (http://heliopropa.irap.omp.eu) and developed during the Planetary Space Weather Services (PSWS) Virtual Activity of the Europlanet H2020 Research Infrastructure funded by the European Union's Horizon 2020 research and innovation programme under grant agreement No 654208 and extended during the Sun Planet Interactions Digital Environment on Request (SPIDER) Virtual Activity of the Europlanet H2024 Research Infrastucture funded by the European Union's Horizon 2020 research and innovation programme under grant agreement No 871149. Work at IRAP is supported by the French National Centre for Scientific Research (CNRS), Centre national d'études spatiales (CNES), and the University of Toulouse III (UPS).

\bibliographystyle{unsrtnat}
\bibliography{references}  






\end{document}